\documentclass[11pt]{article}
\usepackage{amsmath} % AMS Math Package
\usepackage{amsthm} % Theorem Formatting
\usepackage{amssymb}	% Math symbols such as \mathbb
\usepackage{graphicx} % Allows for eps images
\usepackage{multicol} % Allows for multiple columns
\usepackage{pstricks}
\usepackage{authblk}
\usepackage{graphicx}
\usepackage[small]{caption}
\usepackage{subcaption}
\usepackage{float}
\usepackage{amssymb}
\usepackage{bbold}
\usepackage{color}
\usepackage{bbm}
\usepackage{fullpage}
\usepackage{amsmath}
\usepackage{amsthm}
\usepackage{appendix}
\usepackage[normalem]{ulem}

\newcommand{\pif}{P_{i\to f}}

\newcommand{\pd}[2]{\frac{\partial #1}{\partial #2}}
\newcommand{\ket}[1]{\left| #1 \right>} % for Dirac bras
\newcommand{\matrixel}[3]{\left< #1 \vphantom{#2#3} \right|
 #2 \left| #3 \vphantom{#1#2} \right>} % for Dirac matrix elements

\title{Exploring Quantum Control Landscape Structure}
\date{May 19, 2013}
\author[1]{Arun Nanduri}
\author[2]{Ashley Donovan}
\author[2]{Tak-San Ho}
\author[2]{Herschel Rabitz}
\affil[1]{\emph{Department of Physics, Princeton University, Princeton, NJ 08544}}
\affil[2]{\emph{Department of Chemistry, Princeton University, Princeton, NJ 08544}}

\begin{document}
\maketitle

\begin{abstract}
  A common goal of quantum control is to maximize a physical observable through the application of a tailored field. The observable value as a function of the field constitutes a quantum control landscape. Previous works have shown, under specified conditions, that the quantum control landscape should be free of suboptimal critical points. This favorable landscape \textit{topology} is one factor contributing to the efficiency of climbing the landscape. An additional, complementary factor is the landscape \textit{structure}, which constitutes all non-topological features. If the landscape's structure is too complex, then climbs may be forced to take inefficient convoluted routes to finding optimal controls. This paper provides a foundation for understanding control landscape structure by examining the linearity of gradient-based optimization trajectories through the space of control fields. For this assessment, a metric $R\geq 1$ is defined as the ratio of the path length of the optimization trajectory to the Euclidean distance between the initial control field and the resultant optimal control field that takes an observable from the bottom to the top of the landscape. Computational analyses for simple model quantum systems are performed to ascertain the relative abundance of nearly straight control trajectories encountered when optimizing a state-to-state transition probability. The distribution of $R$ values is found to be centered near remarkably low values upon sampling large numbers of randomly chosen initial control fields. Additionally, a stochastic algorithm is used to locate many distinct initial control fields, each of which corresponds to the start of an almost straight control trajectory with $R\simeq 1.0$. The collected results indicate that quantum control landscapes have very simple structural features. The favorable topology and the complementary simple structure of the control landscape provide a basis for understanding the generally observed ease of optimizing a state-to-state transition probability.
\end{abstract}

\section{Introduction}
\begin{figure}[h!]
  \centering
  \includegraphics[width=.8\textwidth]{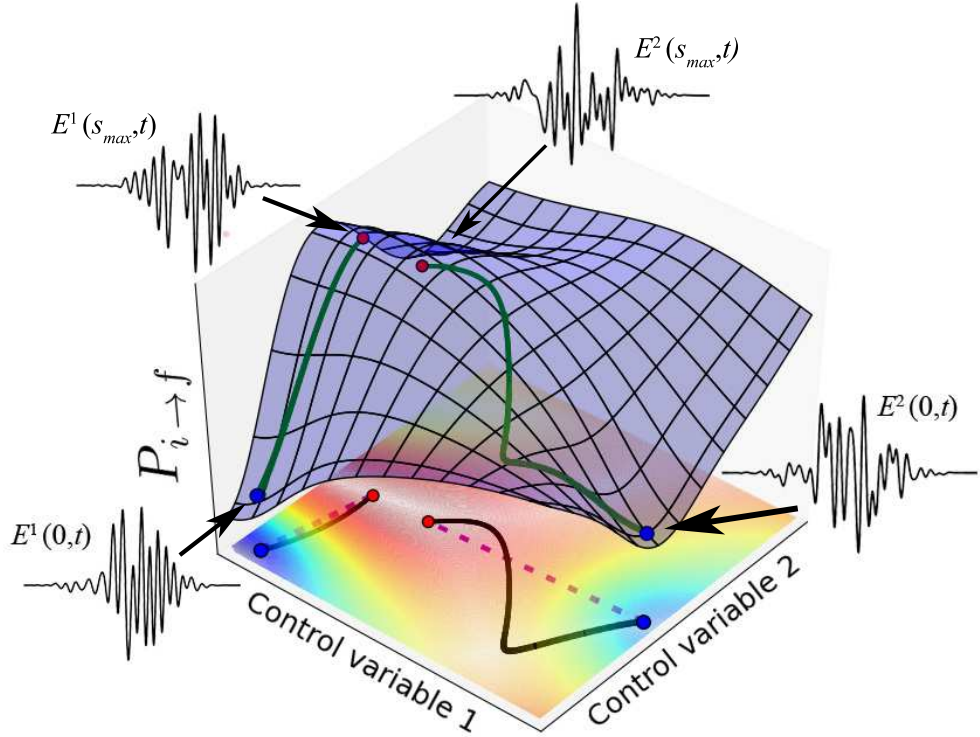}
  \caption[8pt]{(Color online) A schematic illustration of a quantum control landscape (blue surface) as it relates to control space (colored contour map), with two optimization paths shown possessing contrasting $R$ values. The two ordinate axes represent the control variables (in practice there are many more). The abscissa is the observable $\pif$ value on the landscape. Two optimization paths are indicated in green beginning near the bottom of the landscape and ending near the top; they each take the path of steepest ascent. The paths' projections into control space are shown in black; these control trajectories start at some point $E^{\ell}(s=0,t)$, $\forall t$ and end at $E^{\ell}(s=s_{max}, t)$, $\forall t$ for pathways $\ell=1$ or 2. The straight lines connecting the endpoints of the control trajectories are shown as magenta dashed lines. The optimization path $\ell=1$ on the left follows a direct route to the top of the landscape, with the corresponding control trajectory almost overlapping the straight line between its endpoints. Thus, $d_{PL}$ is very close to $d_{EL}$ for this optimization, resulting in $R\sim 1$. Conversely, the optimization path $\ell=2$ on the right follows a convoluted route to the top of the landscape, and its control trajectory takes a circuitous path significantly differing from the straight line between its endpoints. In this case, $d_{PL}$ is much greater than $d_{EL}$, so the optimization along path $\ell=2$ results in a high value of $R$.}
  \label{landscape}
\end{figure}

\begin{figure}[h!]
  \centering
  \includegraphics[width=.8\textwidth]{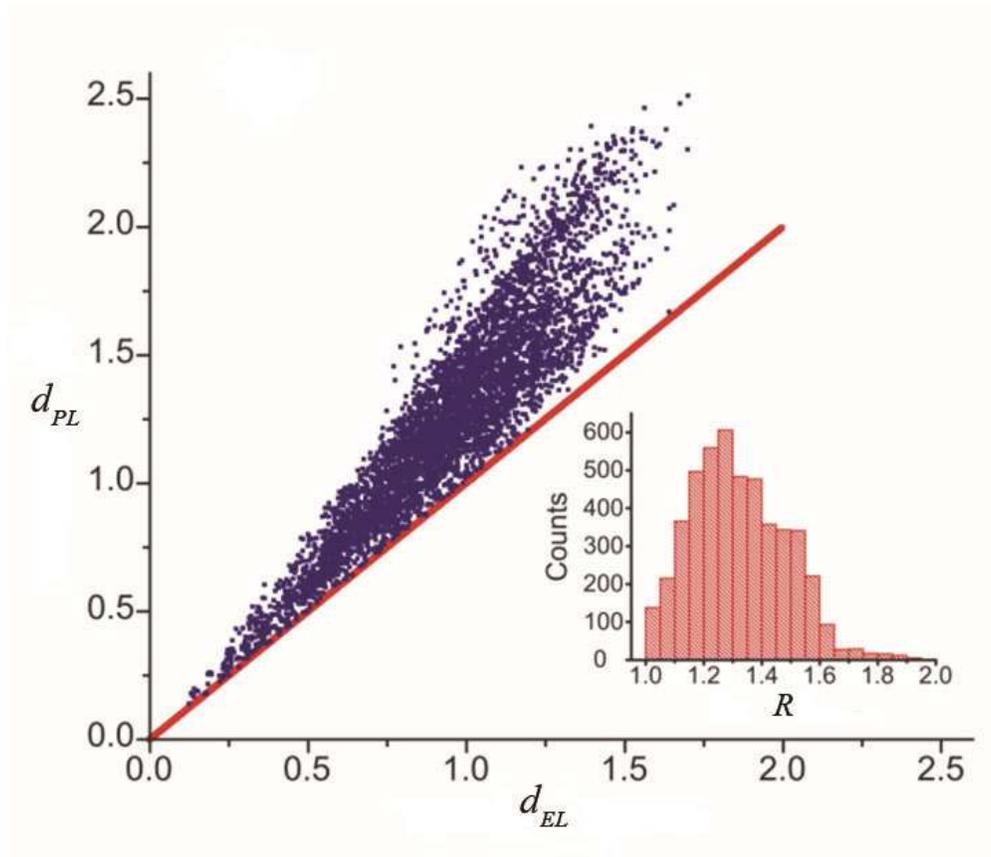}
  \caption{Reproduction of results from Roslund \textit{et al}~\protect\cite{roslund} for an experimental study of second harmonic generation. The scatter plot shows the total length of an optimization trajectory, $d_{PL}$ versus the straight line distance $d_{EL}$ for 4,804 randomly chosen initial fields. The points lying near the red line correspond to $R=\frac{d_{PL}}{d_{EL}}\sim 1.0$. The inset is a histogram for the distribution of the ratio $R$.}
  \label{roslund_figure}
\end{figure}

\indent Advances in femtosecond laser pulse shaping technology \cite{pulseshaping} and the application of closed-loop learning algorithms \cite{lasers} are producing increasing numbers of successful quantum control experiments. In addition, extensive simulations have demonstrated the relative ease of finding optimal control fields~\cite{balint-kurti,W,Moorewithcontrolstuff}. In this paper we consider closed quantum systems in the presence of applied external fields, whose control variables (e.g., phases and/or amplitudes) can be manipulated to achieve high fidelity of a specified physical observable. The dynamics of a quantum system interacting with an applied field is described by the time-dependent Schr\"{o}dinger equation,
\begin{equation}\label{TDSE}
i \hbar \frac{\partial U(t,0)}{\partial t} = H(t) U(t,0), \phantom{space} U(0,0) = \mathbbm{1},
\end{equation}
where $H(t)$ is the Hamiltonian and $U(t,0)$ is the unitary evolution propagator with $0\leq t \leq T$. Using the dipole approximation, the Hamiltonian can be written as
\begin{equation}\label{Ham_dipole}
H(t) = H_0 - \mu E(t),
\end{equation}
where $H_0$ is the field-free Hamiltonian, $\mu$ is the dipole moment and $E(t)$ is the field. \\
\indent The search for a field that produces the highest value for a physical observable at time $T$ can be viewed as occurring on a \textit{quantum control landscape}, which is defined as the observable value as a functional of the field. The observable considered in this work is the pure state transition probability $\pif$ of going from $\ket{i}$ to $\ket{f}$. Optimizing the control field to achieve a high $\pif$ fidelity corresponds to following a path up the quantum control landscape. As a quantum control optimization proceeds, a corresponding trajectory is taken in \textit{control space}, from the initial field producing a low yield to the final field giving the maximum achievable yield. Using a local search algorithm (e.g., gradient ascent) implies that these control trajectories are smoothly varying and can be parametrized by a variable $s\geq 0$. The initial and final control fields are, respectively, $E(s=0,t)$ and $E(s=s_{max},t)$, where $\pif$ has reached a final acceptably high value at $s_{max}$. \\
\indent Under three specified physical assumptions, the landscape is free of suboptimal extrema (points where $\frac{\delta \pif}{\delta E(t)}=0$ $\forall t$ and $0.0<\pif<1.0$)~\cite{Moorewithcontrolstuff,surveypaper,topologyscience,kosut}, whose existence could prematurely halt optimization efforts. These assumptions are that (i) the closed quantum system of $N$ states is controllable, (ii) the resultant set of functions $\frac{\delta U_{ij}(T,0)}{\delta E(t)}$ are linearly independent over $t\in [0,T]$ and (iii) no constraints are placed on the controls. Favorable landscape topology under these assumptions explains in part why excellent and even fully `optimal' yields can routinely be attained in simulations for a variety of physical observables when employing a local search algorithm. However, the \textit{topology} is only one factor in providing a complete description of the landscape features. The \textit{structure} of the landscape is another factor that can significantly influence the complexity of a trajectory taken from $E(0,t)$ through the space of controls to obtain an optimal field $E(s_{max},t)$. Here landscape structure refers to all non-topological (i.e., non-critical point) features. Thus, a full picture of quantum control landscapes entails an understanding of both structural and topological features. Landscape topology has been extensively studied~\cite{Moorewithcontrolstuff,surveypaper,topologyscience,kosut}, and this paper presents an assessment of the companion landscape structure for $\pif$. The present work assumes practical satisfaction of the topological analysis assumptions to permit focusing on the landscape structural features. In this regard, none of the simulations in this work encountered suboptimal extrema. \\
\indent To quantify the linearity of the trajectory taken through the control space during an optimization, we define $R\geq 1$ as the ratio of the control trajectory's total path length $d_{PL}$ to its Euclidean length $d_{EL}$. The path length between the initial and final control fields is given by
\begin{equation}\label{pathlength}
d_{PL} = \int_0^{s_{max}} \left [ \frac{1}{T} \int_0^T \left ( \frac{\partial E(s,t)}{\partial s} \right )^2 dt\right ]^{\frac{1}{2}} \; ds
\end{equation}
and the corresponding Euclidean distance by
\begin{equation}\label{Euclength}
d_{EL} = \left [ \frac{1}{T} \int_0^T \left [ E(s_{max},t) - E(0,t) \right ]^2 dt \right ]^{\frac{1}{2}}
\end{equation}
where $s$ is a variable that tracks position of the field in control space, and $T$ is the target time. The ratio $R$ is then
\begin{equation}\label{ratio}
R = \frac{d_{PL}}{d_{EL}}=\frac{\int_0^{s_{max}} \left [ \int_0^T \left ( \frac{\partial E(s,t)}{\partial s} \right )^2 dt\right ]^{\frac{1}{2}} \; ds}{\left [ \int_0^T \left [ E(s_{max},t) - E(0,t) \right ]^2 dt \right ]^{\frac{1}{2}}}.
\end{equation}
Trajectories that yield $R\simeq 1$ ratios follow direct paths and do not take a gnarled route to achieve optimal control. The extreme limit is an optimal control trajectory that yields $R=1$ constituting a straight path through control space. Consideration of $R$ as a metric of landscape structural features must be linked to an algorithm guiding the trajectory, and in this paper the gradient algorithm is utilized (see Section 2 for details). Fig. \ref{landscape} is a schematic illustrating the relationship between $R$, the quantum control landscape, and the control space. The sketch shows two illustrative paths on the landscape projected into a two dimensional control space (generally the landscape and control space are of high or even infinite dimensionality). The figure caption explains the distinct qualitative difference between the two paths reflected in their $R$ values. \\
\indent In recent experimental work~\cite{roslund}, gradient-based optimizations were performed to maximize second harmonic generation, where the spectral phase expressed in terms of polynomial coefficients acted as the control variables. The ratios $R$ obtained from the experimental results reproduced in Fig. \ref{roslund_figure} indicate that the majority of control trajectories have $R$ values that are surprisingly close to unity (see inset). Simulations utilizing various types of controls with gradient-based optimization routines have monitored the ratio $R$ and also found near $R\simeq 1$ behavior~\cite{complexity,hamiltonianstructure}. The high dimensionality of the quantum control landscape might lead to the intuitive expectation of finding complex landscape structural features. However, the findings of near-straight control trajectories obtained from using gradient ascent algorithms seem to indicate that the controls routinely follow nearly straight paths to achieve optimal yields. This paper will explore this matter further. In addition to being of fundamental interest, knowledge of the structural features along with the landscape topology may open up the prospect of finding better algorithms for performing optimal control. \\
\indent The remainder of the paper is organized as follows. The gradient ascent algorithm utilized to identify control fields that optimize $\pif$ is given in Section 2. Section 3 provides a mathematical analysis of the ratio $R$ and discusses practical upper and lower bounds for $R$. The general form is identified for a control field that would yield a perfectly straight trajectory in control space, for which $R=1$. Section 4 introduces a stochastic algorithm to find \textit{initial} control fields $E(0,t)$ that produce nearly linear control trajectories out to $E(s_{max},t)$ guided by a gradient-based landscape climbing algorithm. The gradient algorithm for \textit{landscape climbing} and the stochastic algorithm for \textit{initial field identification} operate in tandem to respectively maximize $\pif$ and minimize $R$. Section 5 shows the distribution of $R$ values achieved upon starting with a large family of random initial fields. The distribution of distances between members in the family of initial and final fields is then examined, for optimizations with low and high values of $R$. We also verify the mathematical nature of a straight optimization trajectory, formulated in Section 3, by considering a landscape ascent with a small value of $R$. Simulations show that when a value of $R$ corresponds to a truly straight trajectory, continually moving along the direction specified by the \textit{initial} gradient will take a path directly to the top of the landscape. Section 6 provides concluding remarks.

\section{Traversing the quantum control landscape}
\indent The quantum systems considered here have $N$ states such that $H_0$ and $\mu$ (c.f., Eq. (\ref{Ham_dipole})) are $N \times N$ Hermitian matrices. The landscape $\pif [E(t)]$ depends on the control field through the propagator $U(T;0)$ (c.f., Eq. (\ref{TDSE})), where
\begin{equation}\label{Pif}
P_{i \to f} = \vert \langle f \vert U(T;0) \vert i \rangle \vert^2.
\end{equation}
A search for an optimal control field (i.e., a field that yields $P_{i \to f}=1.0$ to specified numerical precision) entails following a trajectory in control space starting at an initial field that gives a $\pif$ value near the bottom of the landscape (an arbitrary field will generally yield $P_{i \to f}\ll 1$) and ending near the top of the landscape. Thus, this process may be viewed as a climb of the associated transition probability control landscape facilitated by manipulation of the field. In this paper, a gradient-based search is employed to find optimal control fields. The choice of the gradient algorithm has the important role of providing a `rule' to follow that specifies the path taken on the landscape and, correspondingly, in control space (c.f., Fig. \ref{landscape}). In particular, the gradient algorithm follows the path of steepest ascent and this makes it the natural rule to follow when considering the subsequent path length ratio $R$ reflecting the landscape structural features. \\
\indent To ensure smooth variations in the controls during a landscape ascent, a parameter $s$ is introduced such that $E(t) \rightarrow E(s,t)$, $s\geq 0$. Monotonically increasing $P_{i \to f}$ as $s$ rises then requires that
\begin{equation}\label{dP_ds}
\frac{dP_{i \to f}}{ds} = \int_0^T \frac{\delta P_{i \to f}}{\delta E(s,t)} \frac{\partial E(s,t)}{\partial s} dt \geq 0, \phantom{space} \forall s\geq 0,
\end{equation}
which can be satisfied by choosing
\begin{equation}\label{deds}
\frac{\partial E(s,t)}{\partial s} = \frac{\delta P_{i \to f}}{\delta E(s,t)},
\end{equation}
where~\cite{DMORPH, otherDMORPH}
\begin{equation}\label{dPdE}
\frac{\delta P_{i\rightarrow f}}{\delta E(t)} = -\frac{2}{\hbar}\Im \bigg \{ \matrixel{i}{U^{\dagger}(T; 0)}{f}\matrixel{f}{U(T; 0)U^{\dagger}(t; 0)\mu U(t; 0)}{i} \bigg  \}.
\end{equation}
This procedure is an implementation of the previously developed D-MORPH algorithm~\cite{DMORPH,otherDMORPH}. In this work, the control variables are the field amplitudes $E(s,t_i)$ at the discretized time points $t_i$, $i=1,2,\dotsc$. Changes in $E(s,t)$ will continue until $\frac{dP_{i \to f}}{ds} = 0$ to acceptable precision at $s=s_{max}$, which occurs at a landscape critical point. Under the three previously specified assumptions, critical points (where $\frac{\delta P_{i \to f}}{\delta E(t)} = 0$, $\forall t$) should only occur at the top and bottom of the landscape~\cite{Moorewithcontrolstuff,surveypaper,topologyscience,kosut}. \\
\indent In all of the simulations to follow, a comparison of the $R$ values for different landscape climbs calls for a common initial observable value $\pif^I$ and a common final observable value $\pif^F$. If an initial trial field gives a yield above (below) $\pif^I$, then D-MORPH is used to climb down (up) the landscape to identify a field $E(s=0,t)$ that produces $\pif^I$. The final value $\pif^F$ is achieved with $E(s=s_{max},t)$ obtained by a D-MORPH climb from $E(s=0,t)$. Thus, each path fully traverses the landscape $\pif^I\rightarrow\pif^F$, corresponding to $E(0,t)\rightarrow E(s_{max},t)$, and the associated $R$ value can then be calculated from Eqs. (\ref{pathlength}-\ref{ratio}).

\section{Quantitative characterization of $R$}
\indent The direct nature of a control space trajectory is reflected by the associated $R$ value in Eq. (\ref{ratio}). The minimum value of $R=1$ corresponds to a perfectly straight control trajectory. While some of the experimental results~\cite{roslund} in Fig. \ref{roslund_figure} and in prior simulations~\cite{complexity,hamiltonianstructure} are within the neighborhood of $R=1$, the majority of optimal solutions yield $R$ ratios somewhat above unity. As a means to understand the general nature of control trajectories that yield $R\simeq 1$, the following analysis formally derives the expected $R=1$ lower bound. The Euclidean distance can be rewritten as
\begin{align}
d_{EL}&=\left [\frac{1}{T}\int_0^T \bigg ( \int_0^{s_{max}} \pd{E(s,t)}{s} ds \bigg )^2 dt\right ]^{\frac{1}{2}} \label{EL_rewrite} \\
&= \left [\frac{1}{T} \int_0^{s_{max}} ds \int_0^{s_{max}} ds' \int_0^T \frac{\partial E(s,t)}{\partial s} \frac{\partial E(s',t)}{\partial s'} dt \right ]^{\frac{1}{2}}. \label{EL_rewrite2}
\end{align}
Using the Cauchy-Schwartz inequality,
\begin{equation}\label{EL_CS}
\int_0^T \frac{\partial E(s,t)}{\partial s} \frac{\partial E(s',t)}{\partial s'}dt \leq \left [ \int_0^T \left ( \frac{\partial E(s,t)}{\partial s} \right )^2 dt\right ]^{\frac{1}{2}} \left [ \int_0^T \left ( \frac{\partial E(s',t)}{\partial s'} \right )^2dt\right ]^{\frac{1}{2}}
\end{equation}
implies that
\begin{align}
d_{EL} &\leq \left [ \frac{1}{T}\int_0^{s_{max}} ds \left [ \int_0^T \left ( \frac{\partial E(s,t)}{\partial s} \right )^2 dt\right ]^{\frac{1}{2}} \int_0^{s_{max}}ds' \left [\int_0^T \left ( \frac{\partial E(s',t)}{\partial s'} \right )^2 dt \right ]^{\frac{1}{2}}\right ]^{\frac{1}{2}} \\
&= \int_0^{s_{max}} ds \left [\frac{1}{T} \int_0^T \left ( \frac{\partial E(s,t)}{\partial s} \right )^2 dt \right ]^{\frac{1}{2}}.
\end{align}
Thus,
\begin{equation}
R \geq \frac{ \int_0^{s_{max}}\left [\int_0^T \Big (\pd{E(s, t)}{s} \Big )^2 \, dt\right ]^{\frac{1}{2}}ds}{ \int_0^{s_{max}} \left [\int_0^T \bigg ( \pd{E(s,t)}{s} \bigg )^2 dt\right ]^{\frac{1}{2}} ds} = 1.
\label{lower bound}
\end{equation}
A stronger result can be obtained when climbing with the gradient algorithm, which sets $\frac{\partial E(s,t)}{\partial s} = \frac{\delta P_{i \to f}}{\delta E(s,t)}$ in Eq. (\ref{deds}). In order to achieve $R=1$, it is sufficient that the gradient function be separable in $s$ and $t$:
\begin{equation}
\label{separable}
\pd{E(s,t)}{s} = \frac{\delta P_{i\rightarrow f}}{\delta E(s,t)} = \alpha (s) \times \beta (t)
\end{equation}
with $\alpha (s)\geq 0$ (to assure a monotonic landscape climb, as explained below). To see that this equation does indeed lead to $R=1$, Eqs. (\ref{separable}) and (\ref{EL_rewrite}) can be substituted into Eq. (\ref{ratio}) as follows:
\begin{align}
\label{rewrittenR}
R &= \frac{\int_0^{s_{max}} \left [\int_0^T\big  (\pd{E(s,t)}{s}\big )^2  dt\right ]^{\frac{1}{2}} ds}{\left [\int_0^T \big ( \int_0^{s_{max}} \pd{E(s,t)}{s} ds \big )^2 dt\right ]^{\frac{1}{2}}} \nonumber \\
&= \frac{\int_0^{s_{max}} \left [\int_0^T \alpha^2(s) \times  \beta^2(t) dt\right ]^{\frac{1}{2}}ds}{\left [\int_0^T \big ( \int_0^{s_{max}} \alpha(s) \times \beta(t) ds \big )^2 dt \right ]^{\frac{1}{2}}} \nonumber \\
&= \frac{\int_0^{s_{max}} \alpha(s) ds \left [\int_0^T \beta^2(t) dt\right ]^{\frac{1}{2}}}{\left [\int_0^T \beta^2(t) dt\right ]^{\frac{1}{2}} \int_0^{s_{max}} \alpha(s) ds} = 1.
\end{align} \\
Integration of Eq. (\ref{separable}) over $s$ and using the initial and final fields $E(0,t)$ and $E(s_{max},t)$, respectively, leads to the identification of $\beta (t)=\frac{E(s_{max},t)-E(0,t)}{\int_0^{s_{max}}\alpha (s)ds}$ and the following practical expression for a straight control trajectory,
\begin{equation}
\label{separable2}
E(s,t) = E(0,t) + \Gamma(s)\Delta E(t)
\end{equation}
where
\begin{equation}
\Gamma(s)=\frac{\int_0^s\alpha(s')ds'}{\int_0^{s_{max}}\alpha(s')ds'}
\end{equation}
and
\begin{equation}
\Delta E(t) = E(s_{max},t) - E(0,t).
\end{equation}
It is apparent that the optimization trajectory in Eq. (\ref{separable2}) interpolates between $E(0,t)$ and $E(s_{max},t)$ on a straight path in control space, with the demand that $\alpha (s) \geq 0$ to assure a monotonic climb of the landscape. The function $\alpha(s)$ controls the `speed' with which the controls proceed along the straight trajectory. \\
\indent An upper bound for $R$ occurs when $d_{PL}$ is maximized and $d_{EL}$ is minimized, in Eqs. (\ref{pathlength}) and (\ref{Euclength}), respectively. If the gradient algorithm is employed, then the expression for $d_{PL}$ can be written in terms of $\frac{\delta P_{i \to f}}{\delta E(s,t)}$,
\begin{align}
d_{PL} &= \int_0^{s_{max}} \left [\frac{1}{T} \int_0^T \left ( \frac{\delta P_{i \to f}}{\delta E(s,t)} \right )^2dt\right ]^{\frac{1}{2}}\; ds \label{plbound} \\
&= \int_0^{s_{max}} \Big \| \frac{\delta P_{i \to f}}{\delta E(s,t)} \Big \|\; ds \nonumber \\
&\leq \int_0^{s_{max}} \frac{2}{\hbar} \| \mu \|\; ds \nonumber \\
&= \frac{2}{\hbar} s_{max} \| \mu \|
\end{align}
where $\| \mu \|$ is the norm of the transition dipole matrix $\mu$~\cite{kosut}. Therefore, the term $d_{PL}$ is bounded from above by the strength of the dipole matrix and the value of $s_{max}$. In practice, the integrand $\left [ \int_0^T \left ( \frac{\delta P_{i \to f}}{\delta E(s,t)} \right )^2dt\right ]^{\frac{1}{2}}$ in Eq. (\ref{plbound}) quickly approaches zero as $s$ increases. We can thus take $s_{max}$ as the value of $s$ where $\Big| \frac{\delta P_{i\to f}}{\delta E(s,t)} \Big |<\epsilon$, $\forall t$ for a specified small tolerance $\epsilon$. Thus, $d_{PL}$ has a practical limit which principally depends on the strength of the dipole. There is no evident avenue to obtaining an analytical lower bound on $d_{EL}$, although very small values of $d_{EL}$ likely correspond to undesirable non-robust initial and final control fields.

\section{Searching for fields to minimize $R$}
\indent A prime goal of this paper is to assess the frequency of finding control trajectories at very small $R$ values. The D-MORPH gradient method in Section 2 aims to go from an initial field $E(0,t)$ to a final field $E(s_{max},t)$ that produces a high state-to-state transition probability. This procedure makes no attempt to address the value of $R$. In order to minimize $R$ and identify fields that follow an especially direct path in control space, a stochastic Particle Swarm Optimization (PSO) algorithm~\cite{PSO}, is employed along with the D-MORPH optimization method; further details on the PSO algorithm are presented in an Appendix, and a brief overview of the algorithm is given below. \\
\indent Each particle, or field $E_k(t)$, $k=1, 2, \dotsc, K$ in a set of $K$ trial control fields is `normalized' to start at the same initial $P_{i\rightarrow f}^I$ value as explained in Section 2. An optimization using D-MORPH is then performed to reach the final $\pif^F$ value for each of the $K$ trial fields, and the corresponding $R$ values are recorded for all $K$ optimization trajectories. The initial field that produced the lowest value of $R$ upon D-MORPH optimization, denoted by $E^{best}_{swarm}(t)$, is then treated as the best field and is used as a seed to update the $K$ trial fields. These new $K$ fields form the next trial set, and $E^{best}_{swarm}(t)$ is updated with the new best field. The new $E^{best}_{swarm}(t)$, as well as the best field in each particle's own history, $E^{best}_k$, are used to again update the set of trial control fields. The algorithm then continues for a predetermined number of generations, ideally approaching $R\simeq 1$. \\
\indent The D-MORPH landscape climbing algorithm makes use of the field at a large number of time points as the control variables to assure an accurate landscape climb and avoid introducing artificial landscape features. However, the PSO algorithm searching for new trial control fields utilizes a smaller number of frequency components of the field (c.f., Eq. (\ref{field_parametrized})) as the set of initial control variables, thus placing a practical limit on the freedom in seeking to minimize $R$. Using parametrized initial fields with a reduced number of variables was necessary to accelerate the PSO algorithm. Nevertheless, we will show that $R$ can be dramatically reduced through employing the gradient and PSO algorithms in tandem. Further minimization of $R$ may require more flexible control fields, and possibly a better algorithm than PSO for seeking the best initial field. A basic open question is whether attaining $R=1$ is in principle possible under particular physical conditions.

\section{Numerical Illustrations}
\indent The following illustrations aim to assess the attainable values for $R$ over a large body of simulations. Section 5.1 considers the distribution of $R$ values obtained by starting with a large set of randomly chosen initial fields. The results of these simulations lay the foundation for the assessment in Section 5.2 that $R\lesssim 1.05$ can be considered as characterizing a nearly straight trajectory. In particular, various members of this set whose $R$ values range from low to high are tested in Section 5.2 for the ability of the gradient evaluated at just the \textit{initial} field to successfully specify a `straight shot' up the landscape. The stochastic PSO algorithm is used in Section 5.3 to search for initial control fields that yield minimal (near-unity) values for $R$. For clarity of presentation, the Hamiltonians illustrated here are all of dimension $N=5$. Additional simulations (not shown) and earlier preliminary work~\cite{complexity,hamiltonianstructure} indicate that the value of $N$ does not significantly influence the value of $R$. All of the variables in the simulations are expressed in dimensionless units.

\subsection{Statistical behavior of $R$}
\begin{figure}[h!]
  \centering
  \includegraphics[width=1\textwidth]{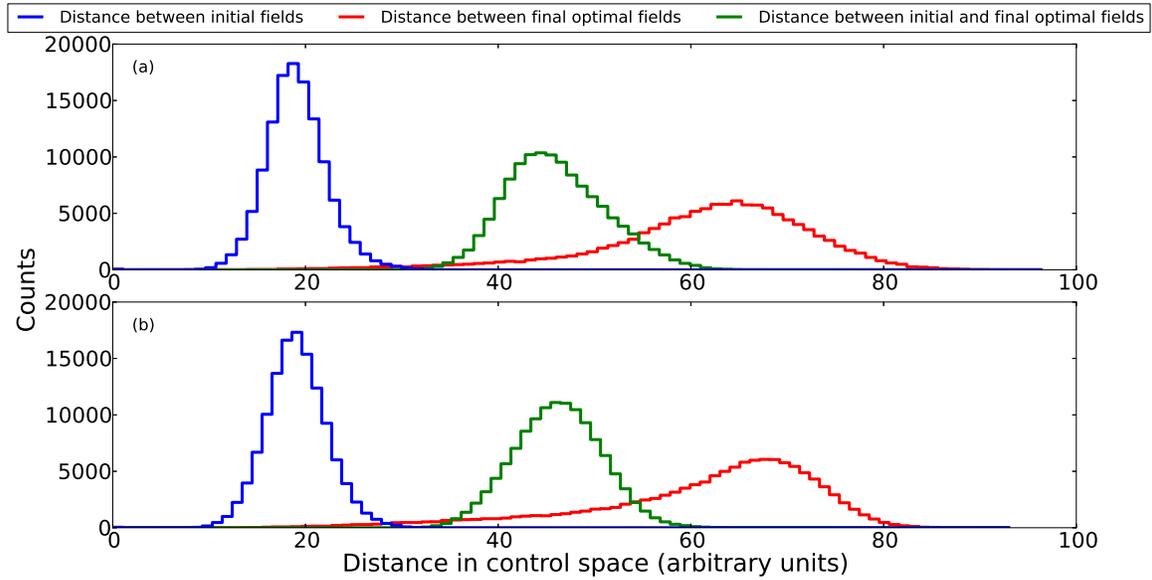}
  \caption{(Color online) Distributions of pairwise distances between initial control fields, final control fields, and initial-final field pairs. (a) The distance distributions resulting from the 500 optimization trajectories that yielded the lowest values of $R$ with an average of 1.071 drawn from the 2,000 optimizations performed using the Hamiltonian in Eqs. (\ref{H0_simulations}) and (\ref{dipole_simulations}). (b) The distance distributions resulting from the 500 optimization trajectories that yielded the highest values of $R$, with an average $R$ value of 1.175. The similarity of the two plots suggests that control trajectories with low $R$ values are distributed throughout control space in the same way as control trajectories with high $R$ values.}
  \label{fig:splithist}
\end{figure}

\begin{figure}[h!]
  \begin{center}
  \includegraphics[width=0.8\textwidth]{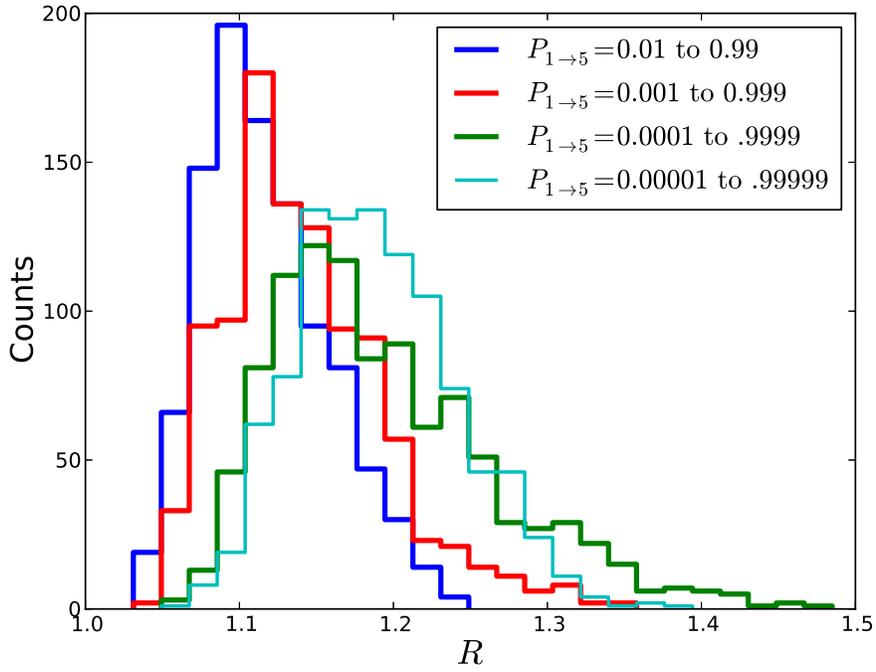}
  \caption{(Color online) The distribution of $R$ values resulting from increasing the demand on the fidelity of $P_{1\to 5}$ at the beginning and end of the landscape climbs. 1,000 runs were performed for each demanded level of transition probability accuracy. A systematic slight increase in $R$ values is evident as the climbs stretch to further extremes of the landscape, but the effect is small.}
  \label{fig:R_Pif_demands}
  \end{center}
\end{figure}

\indent As a means to assess the complexity of the control landscape structural features, 2,000 individual state-to-state transition probability optimizations were performed, followed by computation of the resulting $R$ values. The optimizations used a Hamiltonian with
\begin{equation}\label{H0_simulations}
H_0=\begin{pmatrix} -10 & 0 & 0 & 0 & 0 \\ 0 & -7 & 0 & 0 & 0 \\ 0 & 0 & -3 & 0 & 0 \\ 0 & 0 & 0 & 2 & 0 \\ 0 & 0 & 0 & 0 & 8
\end{pmatrix}
\end{equation}
and
\begin{equation}\label{dipole_simulations}
\mu=\begin{pmatrix} 0 & \pm 1 & \pm 0.5 & \pm 0.5^2 & \pm 0.5^3 \\ \pm 1 & 0 & \pm 1 & \pm 0.5 & \pm 0.5^2 \\ \pm 0.5 & \pm 1 & 0 & \pm 1 & \pm 0.5 \\ \pm 0.5^2 & \pm 0.5 & \pm 1 & 0 & \pm 1 \\ \pm 0.5^3 & \pm 0.5^2 & \pm 0.5 & \pm 1 & 0
\end{pmatrix},
\end{equation}
where the signs of the dipole matrix elements were chosen randomly once (subject to the constraint that $\mu$ be symmetric) and then fixed for each of the 2,000 runs. The target observable was $P_{1 \to 5}$, and the initial control field was parametrized with the form
\begin{equation}\label{field_parametrized}
E(t)=\frac{1}{F} \mbox{exp}[-0.3 (t-\frac{T}{2})^2] \sum_{n=1}^{20} a_{n} \, \mbox{sin}(\omega_n t + \phi_{\omega}),
\end{equation}
where the final time is $T=10$. The form in Eq. (\ref{field_parametrized}) is only used to specify the initial field, which was then discretized into 1,001 time points for optimization. The field frequency components, $\omega_n=n$, permit substantial resonant and near resonant interaction with the system's internal energy level spacings (described by $H_0$). The amplitudes $a_n$ and phases $\phi_n$ were randomly chosen respectively from the uniform distributions [0,1] and $[0,2\pi]$. The normalization factor $F$ is set such that the maximum absolute value of the initial field at any time never exceeds 1, which is done to limit the strength of the field. In these simulations the trajectories were normalized to initial and final $P_{1\to 5}$ values of $P_{1\to 5}^I=0.01$ and $P_{1\to 5}^F=0.99$, respectively. Over the course of the optimizations carried out in this paper, the fluences of the initial control fields typically rose slightly but maintained the same order of magnitude. Similarly, the optimal control fields did not pick up additional frequency structure compared to the trial control fields obtained from Eq. (\ref{field_parametrized}). \\
\indent Figure \ref{R_hist} shows the behavior of the computed $R$ values from the 2,000 simulations, where a histogram distribution of $\mbox{log}(R-1)$ is presented to reveal details (the inset plot gives the distribution of $R$ on a linear scale). Out of all the simulations, the smallest $R$ value was 1.02, the average was 1.12, and the largest was 1.26. The distribution is skewed towards small $R$ values, implying that the family of landscape climbs did not encounter highly gnarled landscape structural features. A quantitative means for interpreting values of $R$, such as 1.02 and 1.26, will be addressed in Section 5.2 after the statistical assessments below. \\
\indent The dipole matrix $\mu$ in Eq. (\ref{dipole_simulations}) permitted the control field to access any $\vert j \rangle \to \vert k \rangle, j \neq k$ transition, although with diminished strength for increasing values of $|j-k|$. To determine if the behavior of $R$ is strongly dependent upon the allowed state transitions, the following two dipole matrices were also used in separate optimizations over 1,000 random trial fields. A less restrictive dipole is described by
\begin{equation}\label{dipole_free}
\mu=\begin{pmatrix} 0 & \pm 1 & \pm 1 & \pm 1 & \pm 1 \\ \pm 1 & 0 & \pm 1 & \pm 1 & \pm 1 \\ \pm 1 & \pm 1 & 0 & \pm 1 & \pm 1 \\ \pm 1 & \pm 1 & \pm 1 & 0 & \pm 1 \\ \pm 1 & \pm 1 & \pm 1 & \pm 1 & 0
\end{pmatrix}
\end{equation}
and a more restrictive dipole is given by
\begin{equation}\label{dipole_restrict}
\mu=\begin{pmatrix} 0 & \pm 1 & 0 & 0 & 0 \\ \pm 1 & 0 & \pm 1 & 0 & 0 \\ 0 & \pm 1 & 0 & \pm 1 & 0 \\ 0 & 0 & \pm 1 & 0 & \pm 1 \\ 0 & 0 & 0 & \pm 1 & 0
\end{pmatrix},
\end{equation}
both with random signs in keeping with $\mu$ being symmetric. Upon using the dipole of Eq. (\ref{dipole_free}), the minimum, average, and maximum $R$ values were 1.005, 1.05, and 1.17, while for the dipole of Eq. (\ref{dipole_restrict}), the minimum, average, and maximum $R$ values were 1.07, 1.24, and 1.64, respectively. Comparing these statistical measures while considering the dipoles in Eq. (\ref{dipole_simulations}), (\ref{dipole_free}), and (\ref{dipole_restrict}), it is evident that less restrictive dipole operators skew the distribution of $R$ values more toward straight trajectories through the control space. The variations in $R$ with respect to Hamiltonian structure opens up the question of whether particular control mechanisms are correlated to the value of $R$. This issue is a complex matter to understand, as $R$ reflects the collective dynamics along a complete climb of the landscape for $\pif^I\leq \pif \leq \pif^F$. As an initial glimpse at this topic, the population dynamics $P_{1\to j} (t)$, $j=1,\ldots,5$ was examined at the final field $E(s_{max},t)$ for several $R$ values with the dipoles in Eqs. (\ref{dipole_simulations}), (\ref{dipole_free}), and (\ref{dipole_restrict}). Such population plots are commonly reported in the literature~\cite{surveypaper,popdynam1,photonicreagent}, and those examined here were quite varied in mechanistic character (results not shown) with no clear evident relation to the value of $R$. A full analysis of control mechanism in relation to $R$ is beyond the scope of the present paper. \\
\indent We now consider whether the control trajectories that result in low $R$ values correspond to fields that are distributed in control space differently than those that result in high $R$ values. As a means to examine this issue, the pairwise Euclidean distances were calculated between all pairs of (1) initial fields, (2) final optimal fields, and (3) initial-final fields using the analog of Eq. (\ref{Euclength}):
\begin{equation}
d_{EL}^{ij} = \left [ \frac{1}{T}\int_0^T [E_i(t)-E_j(t)]^2 dt \right ]^{\frac{1}{2}}
\end{equation}
for the $i$th and $j$th pair of fields. From the 2,000 simulations described above that used the dipole of Eq. (\ref{dipole_simulations}), the distance distributions of the control trajectories from the 500 simulations with the lowest $R$ values (average $R = 1.071$) are shown in Fig. \ref{fig:splithist}(a). The distribution of distances between initial fields (blue curve) is shifted to considerably smaller values and with a narrower variance than those corresponding to the final optimal fields (red curve), which implies that optimal fields are spread more widely across control space than the initial fields. The green curve corresponds to the pairwise distances between all initial and optimal fields, whose distribution lies midway between the latter two. The same distance distributions were computed for the 500 control trajectories associated with the highest $R$ values with an average of $R=1.175$. The results are shown in Fig. \ref{fig:splithist}(b) and are virtually indistinguishable from the distributions in Fig. \ref{fig:splithist}(a). Importantly, this behavior implies that fields corresponding to low $R$ trajectories are distributed in a similar manner as those corresponding to higher $R$ trajectories. Consequently, low $R$ trajectories are as prevalent as relatively high $R$ trajectories, which provides a basis to understand why near-straight control trajectories do not appear to be difficult to find~\cite{roslund,complexity,hamiltonianstructure}. \\
\indent In section 4, an upper bound for the path length $d_{PL}$ was shown to depend on the value for $s_{max}$. In principle, starting from the absolute bottom of the landscape at $\pif=0$ and then achieving perfect control where $P_{i \to f} = 1$ would require $s_{max} \to \infty$. In this regard, we considered the effect of increasing the required degree of control upon $R$ with the following four optimization scenarios: (1) $P_{1\to 5}^I=0.01 \to P_{1\to 5}^F=0.99$, (2) $P_{1 \to 5}^I = 0.001 \to P_{1 \to 5}^F = 0.999$, (3) $P_{1 \to 5}^I = 0.0001 \to P_{1 \to 5}^F = 0.9999$, and (4) $P_{1 \to 5}^I = 0.00001 \to P_{1 \to 5}^F = 0.99999$. For each case, 1,000 D-MORPH simulations were performed using the Hamiltonian of Eqs. (\ref{H0_simulations}) and (\ref{dipole_simulations}) with random initial fields, and the resulting $R$ values were computed. Fig. \ref{fig:R_Pif_demands} shows the distribution of $R$ values for the four scenarios. The corresponding average values of $R$ were:  1.117 for case (1); 1.142 for case (2); 1.194 for case (3); and 1.190 for case (4). This trend is consistent with the intuitive expectation that demanding more extreme values for the start and end of an optimization landscape encounters additional landscape structural features. However, the effect is very  modest considering the increase of three orders of magnitude in precision for $P^I_{1\to 5}$ and $P^F_{1\to 5}$.

\subsection{Assessment of `straight shot' character through control space}
\begin{figure}[h!]
  \centering
  \includegraphics[width=1\textwidth]{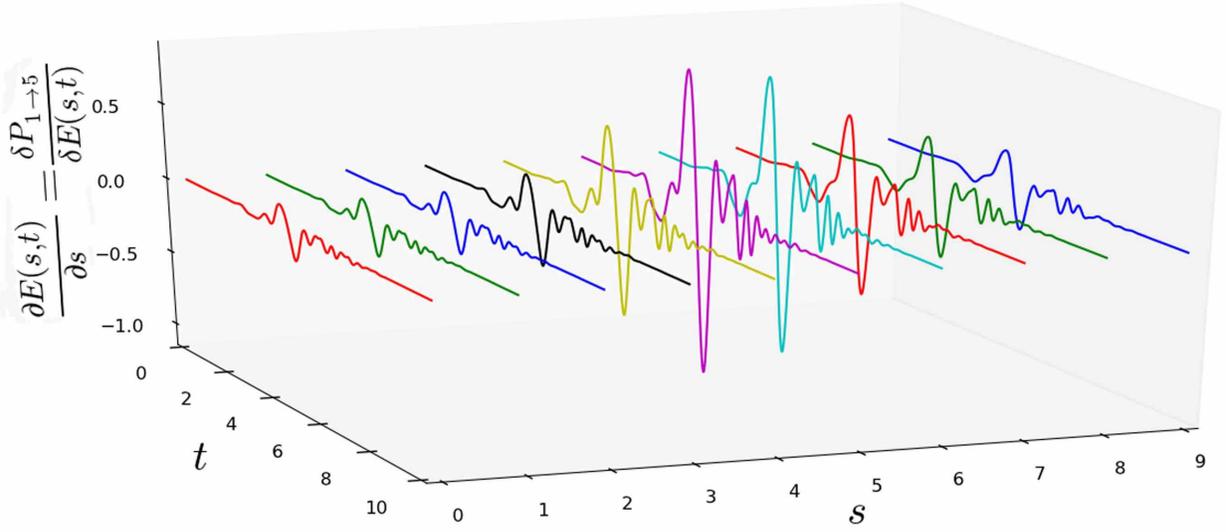}
  \caption{(Color online) A succession of plots of the gradient $\frac{\delta P_{1\to 5}}{\delta E(s,t)}$ over time $t$ as a control trajectory with a low value of $R=1.0036$ proceeds over $s$ towards an optimal field. From Eq. (\ref{separable2}), the cross-section at each value of $s$ should be a scaled version of the function $\Delta E(t)$. From the figure, these cross sections each have essentially the same time dependence, as expected. The overall amplitude as $s$ varies is modulated by the function $\alpha (s)\geq 0$, whose bell-shaped character can be discerned from the figure.}
  \label{straightline}
\end{figure}

\begin{figure}[h!]
  \centering
  \includegraphics[width=.8\textwidth]{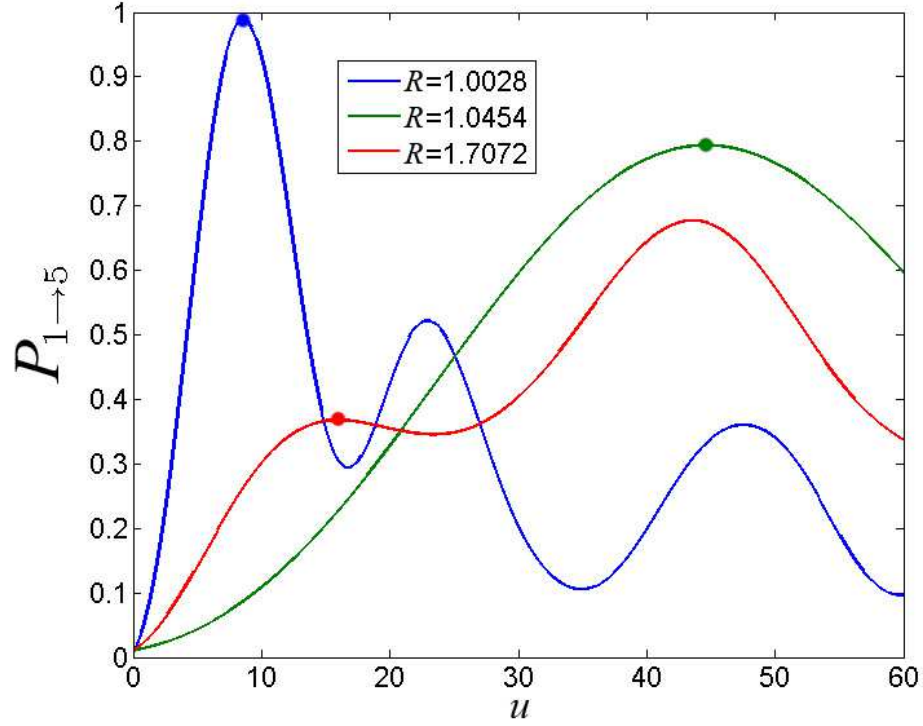}
  \caption{(Color online) The value of $P_{1\to 5}$ obtained by traversing control space upon continuously marching in the direction specified by the initial gradient $\frac{\delta P_{1\to 5}}{\delta E(s=0,t)}$ of the control field using Eq. (\ref{straightshoteq}). Three trials are shown, each corresponding to a different value of $R$. The value of $P_{1\to 5}$ at its first local maximum (indicated by the dot on each curve) is the measure of success achieved by continuing to move with increasing $u$ in the same direction specified by $\frac{\delta P_{1\to 5}}{\delta E(s=0,t)}$. The yield obtained in this way depends significantly on $R$, with $R\lesssim1.05$ corresponding to successful straight shots through control space.}
  \label{fig:straightshot}
\end{figure}

\indent As a first assessment of the straight shot trajectory behavior set out in Section 3, we examined a climb with a very low $R$ value of 1.0036 using the Hamiltonian of Eqs. (\ref{H0_simulations}) and (\ref{dipole_simulations}) along with starting and ending points of $P_{1\to 5}^I=0.01$ and $P_{1\to 5}^F=0.99$. This climb was obtained by minimizing $R$ utilizing the PSO algorithm described in Section 4 and later in Section 5.3. To verify the behavior of $E(s,t)$ predicted by Eq. (\ref{separable}), Fig. \ref{straightline} shows $\frac{\delta P_{i\rightarrow f}}{E(s,t)} = \pd{E(s,t)}{s}$ as a function of $t$ along the optimization trajectory over $s$. No attempt was made to fit the data to the form in Eq. (\ref{separable}), but Fig. \ref{straightline} displays the expected feature of a single function of time $\beta (t)$ modulated by a positive time-independent function $\alpha(s)$. The behavior of $\alpha(s)$ peaking in the middle of the climb and becoming small at either end is in accord with observations that a gradient-based optimization proceeds much more quickly in the middle than at the beginning and end of the climb. \\
\indent The collected simulation results in Section 5.1 using random initial fields produced $R$ values ranging from 1.005 to 1.7. An important goal is to have a quantitative sense of what values of $R$ can be considered as small and as indicative of a true straight shot across the control space. This assessment can be performed by using the initial gradient $\frac{\delta P_{i\to f}^I}{\delta E(s=0,t)}=\pd{E(s,t)}{s} \Big |_{s=0}$ to point in a direction in the control space, or equivalently on the landscape, and then continuing to move in that same direction to determine the maximum attained $\pif$ value. The following relation mathematically expresses this simple rule for traversing the control landscape
\begin{equation}
E(u,t)=\left (\frac{\delta P_{i\to f}}{\delta E(s=0,t)}\right )u + E(0,t), \; \; u\geq 0.
\label{straightshoteq}
\end{equation}
Equation (\ref{straightshoteq}) has the same form as Eq. (\ref{separable}) upon integration over $s$, with $\beta(t)$ chosen as $\frac{\delta P_{i\to f}}{\delta E(s=0,t)}$ and $u$ serving the role of $\int_0^s\alpha(s')ds'$ such that Eq. (\ref{straightshoteq}) dictates a procedure that marches in the same initial direction in control space specified by the gradient. This equation was tested by drawing on particular runs with distinct $R$ values from the 2000 simulations and one PSO minimization of $R$ using the dipole function in Eq. (\ref{dipole_simulations}). Fig. \ref{fig:straightshot} shows the $P_{1 \to 5}$ trajectories from three different simulations using Eq. (\ref{straightshoteq}) from initial fields producing low, medium, and high $R$ values. The measure of straight trajectory character is reflected by the first $P_{1\to 5}$ maximum encountered, shown as a dot on the curve in each case. For the case $R=1.0028$ (obtained with the PSO algorithm), the maximum of $P_{1 \to 5}$ was 0.9888, which is very close to the value $P_{1 \to 5} = 0.99$ obtained from the D-MORPH climb using the same initial field. With $R=1.0454$, the maximum was $P_{1 \to 5}=0.7940$, and finally the initial field corresponding to $R=1.7072$ produced $P_{1 \to 5}=0.3697$. In assessing these results, an important point is that the D-MORPH optimization starting with the associated initial field in each case produced a trajectory that ended up at the top of the landscape, taken as $P_{1\to 5}^F=0.99$. Thus, from Fig. \ref{fig:straightshot} and other similar tests (not shown) we may consider $R\lesssim 1.05$ as `small' such that the trajectory may be characterized as a straight shot giving a high yield of $P_{i\to f}\widetilde{>}0.8$. We emphasize that this threshold of $R\lesssim 1.05$ is based on the very demanding criterion that the initial ascent direction in Eq. (\ref{straightshoteq}) has global significance over the full control space trajectory. Thus, even a trajectory with $R\sim 2$ could be considered as not encountering complex landscape structural features, yet also not fully straight by the unsuccessful application of Eq. (\ref{straightshoteq}). \\
\indent Besides using Eq. (\ref{straightshoteq}) as a basis to assess the meaning of a straight shot, an intriguing prospect would be the conversion of Eq. (\ref{straightshoteq}) into an efficient high yield control landscape climbing algorithm. This prospect requires \textit{a priori} knowledge about the location of a trial field that subsequently would give a low $R$ value. The distribution in Fig. \ref{fig:splithist} implies that such good trial fields are broadly scattered about control space. Unfortunately, the same figure also indicates that less attractive trial fields, from a straight shot perspective, are also widely dispersed in the control space. The exploitation of Eq. (\ref{straightshoteq}) as a climbing algorithm remains an open challenge. Notwithstanding this comment, the distribution of $R$ values for all of the D-MORPH gradient climbs are nearly straight from the results in Figs. \ref{R_hist} and \ref{fig:R_Pif_demands}, as well as in the prior literature~\cite{complexity,hamiltonianstructure} and the experiments shown in Fig. \ref{roslund_figure}.

\subsection{Optimizing $R$}
\begin{figure}[h!]
  \centering
  \includegraphics[width=.8\textwidth]{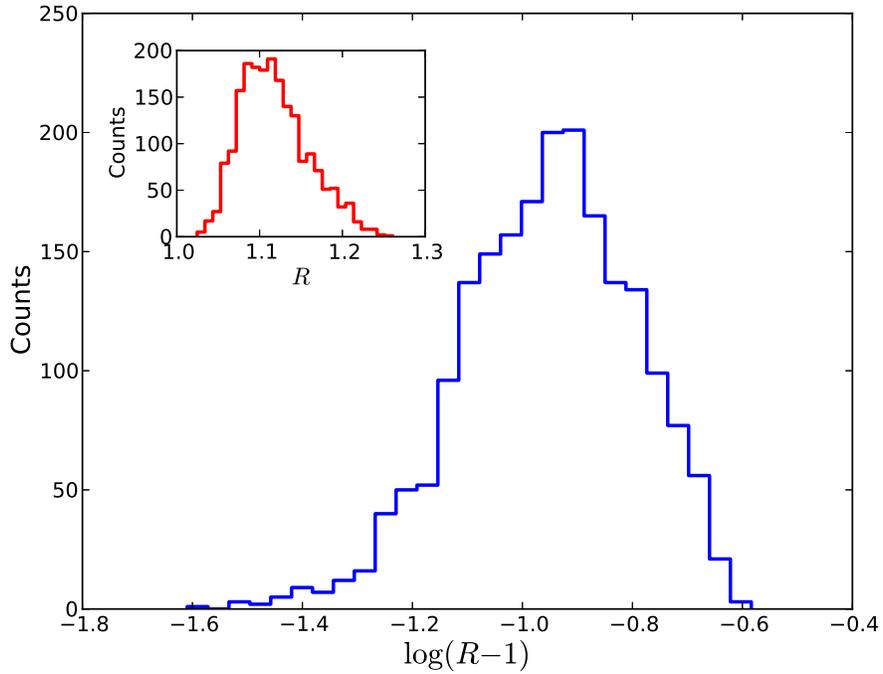}
  \caption{(Color online) The distribution of $\mbox{log}(R-1)$ values for 2,000 $P_{1 \to 5}$ optimizations performed using random initial fields with the Hamiltonian in Eqs. (\ref{H0_simulations}) and (\ref{dipole_simulations}). $R$ values were computed for each optimization trajectory where the starting and final values are, respectively, $P_{1 \to 5}^I = 0.01$ and $P_{1 \to 5}^F = 0.99$. The average $R$ value from the 2,000 optimizations was 1.12. The inset shows the $R$ values on a linear scale, indicating the tendency to attractively cluster at smaller values.}
  \label{R_hist}
\end{figure}

\indent Section 4 described the stochastic PSO algorithm to search for initial control fields that lead to $R\to 1.0$. The procedure was implemented for the Hamiltonian in Eqs. (\ref{H0_simulations}) and (\ref{dipole_simulations}) with $K=50$ initial fields randomly chosen through selecting 20 amplitudes and 20 phases in Eq. (\ref{field_parametrized}). After 50 generations of the PSO procedure, $R-1$ was driven down to $\sim 10^{-4}$ across multiple deployments of the algorithm. This result is two orders of magnitude lower than $R=1.02$, which was the smallest value found from the 2,000 random initial fields in Section 5.1. Application of the PSO algorithm demonstrates that there are locations in control space where very straight paths can be taken to the top of the landscape. At these particular locations, the landscape all the way to the top is especially smooth and devoid of confounding structures, compared to other regions which possess some degree of gnarled landscape structure, as indicated by the results of Figs. (\ref{R_hist}) and (\ref{fig:R_Pif_demands}). However, we emphasize that from an absolute perspective, all the domains of the landscape explored here and sampled in prior work~\cite{complexity,hamiltonianstructure} appear to have smooth structural features, reflected in the observed values of $R$ being less than $\sim 2$. \\
\indent While we naturally used the PSO algorithm to seek initial fields that produced the \textit{lowest} $R$ values, it is possible to redirect the PSO algorithm to search for initial fields that have the \textit{highest} $R$ values. This goal was pursued as a means to determine an upper limit on $R$, with regard to the discussion in Section 3. Using the same PSO algorithm parameters utilized above, three separate PSO maximizations of $R$ were performed. The resulting values for $R$ were found to be 1.702, 1.608, and 1.718, which are noticeably larger than the maximum value of $R\approx 1.3$ from random initial fields in Fig. \ref{R_hist}. The similarity of the three extreme $R$ values from the PSO maximization procedure hints at a practical upper bound on $R$. Once again, these worst case $R$ values are themselves not large, indicating the minimal presence of coarse landscape structural features.

\section{Conclusions}
\indent This paper takes a first step towards systematically investigating quantum control landscape structural features beyond topological critical point features. The research was motivated by results from~\cite{roslund} indicating a surprising degree of straight shot character taken during quantum control experiments, as well as simulations exhibiting similar behavior under various control scenarios~\cite{complexity,hamiltonianstructure}. The numerical illustrations in this paper consistently showed that the path length ratio $R$ is modestly above 1.0, even for trajectories starting from randomly chosen initial fields. We emphasize that all the trajectories studied in this work are the result of using a gradient-based landscape climbing algorithm, whose adherence to the path of steepest ascent makes it a natural choice to explore the landscape structure. For comparison, upon using a stochastic climbing algorithm, we found values of $R\gg 1$ by constructing a trajectory based on the location of the best member of the population in each generation. \\
\indent The PSO algorithm was effective in minimizing $R$ down to $R-1\sim 10^{-4}$. These results leave open a controllability-like question of under what physical conditions it is possible to reach $R=1.0$ using some initial control and specified values of $P_{i\to f}^I$ and $P_{i\to f}^F$. A related open issue is whether an identified initial field is isolated or a member of a manifold that gives a specific $R$ value, especially approaching $R\to 1$. The finding that there may exist precisely straight control trajectories, which requires that the gradient $\frac{\delta P_{i \to f}}{\delta E(s,t)}$ be separable into an $s$-dependent and a $t$-dependent term in Eq. (\ref{separable}), is intriguing, since ostensibly the gradient is a highly complex functional of $E(s,t)$ through the propagator $U(t,0)$ (c.f., Eq. (\ref{dPdE})). Future work~\cite{Nanduri2} will further explore this result, in addition to examining the landscape structure for optimizing arbitrary physical observables and unitary transformation objectives. More numerical simulations, possibly with new algorithms, on wider classes of systems would be valuable as empirical evidence to guide mathematical analyses of landscape structure. In this regard, a prime practical topic is establishing how control constraints conspire to introduce artificial landscape features, and eventually traps~\cite{constrained}. Similarly, significant control constraints can limit reaching the lowest values of $R$ by forcing trajectories to take circuitous routes. \\
\indent Quantum control experiments often utilize hundreds of control variables in an attempt to extremize a control objective. The high dimensionality of the control search spaces nominally would seem to introduce complex structure into the corresponding control landscapes, but the identification of gradient ascents from arbitrary initial control fields being almost straight indicates that the landscape structure is far simpler. These results complement research showing that attractive landscape topology generally exists under specified reasonable physical assumptions~\cite{surveypaper,Moorewithcontrolstuff,topologyscience,kosut}. Taken together, the dual salutary landscape features provide a more comprehensive basis to understand the repeated ease and efficiency of finding good controls. This paper sets the foundation for further investigations of control landscape structure and the associated trajectories through control space.

\newpage
\appendixtitleon
\begin{appendices}

\section*{Appendix: Summary of the particle swarm optimization\\ algorithm}
\indent A particle swarm optimization (PSO) algorithm searches for the solution to an optimization problem using a stochastic approach that mimics an evolutionary strategy~\cite{PSO}. In the present work, the PSO algorithm searches for an initial control field that (after a D-MORPH-assisted $P_{i \to f}$ optimization) yields a minimal value for the associated control trajectory's path length ratio $R$. The PSO algorithm begins by creating a set of $K$ trial control fields (considered as the swarm of candidate solutions), where each field is referred to as a particle. In the present work, $K=50$, and the control variables accessible to the PSO algorithm are the field amplitudes and phases from the domains [0,1] and [0,2$\pi$], respectively (c.f., Eq. (\ref{field_parametrized})). Using the Hamiltonian of Eqs. (\ref{H0_simulations}) and (\ref{dipole_simulations}), each of these trial fields was then normalized to yield the initial value $P_{1 \to 5}^I = 0.01$ using the D-MORPH optimization technique over a grid of 1,001 time points. The initial fields are then optimized to reach $P_{1 \to 5}^F = 0.99$ again using the D-MORPH optimization technique, and $R$ is computed for each of the 50 optimizations. The initial field corresponding to the trajectory with the lowest $R$ value is denoted by $E^{best,g}_{swarm}(t)$, where $g$ represents the generation index. In the present work, the maximum number of generations was set at 50. Using the location of $E^{best, g}_{swarm}(t)$ as well as the location of the best initial field encountered by each individual particle $E_k^{best,g}$, $k=1, 2, \dotsc, 50$, the velocity, or change in each particle's position, is then computed as follows~\cite{PSO}
\begin{equation}\label{velocity}
v_k^{g} = C_0v_k^{g-1} + C_1S_1(E_{swarm}^{best,g-1} - E_{k}^{best,g-1}) + C_2S_2(E_{swarm}^{best,g-1}-E_{k}^{g-1}),
\end{equation}
which contains the following parameters: $C_0$, inertia (defined below), $C_1$, the cognitive attraction, which we set to 0.5, and $C_2$, the social attraction, which we set to 1.5. The diagonal matrices $S_1$ and $S_2$ have entries that are randomly chosen to be 0 or 1 and are the source of the stochasticity in the PSO algorithm. The inertia $C_0$ is given by~\cite{PSO}
\begin{equation}\label{inertia}
C_0 = 0.9-\frac{g-1}{150}.
\end{equation}
As $g$ increases, $C_0$ decreases, causing the particles to coalesce around a solution. The updated positions of the particles are set through
\begin{equation}\label{update_particle}
E_k^{g} = E_k^{g-1} + v_k^{g}.
\end{equation}
\indent The PSO algorithm is not guaranteed to find the global minimum of $R$ because of the possibility that the swarm may collapse around a local solution; this outcome can happen when each particle in the swarm converges on the same initial control field and hence the same $R$ value, at the expense of exploring more optimal minima in $R$. This phenomenon may be problematic, as there is no available analysis of the functional relationship between $R$ and the control variables (i.e., the $R$ landscape). In particular, we do not know if there exist local minima for $R$, with a given $P^I_{i\to f}$ and $P^F_{i\to f}$, that could cause the PSO swarm to converge prematurely.

\end{appendices}

\section*{Acknowledgments}
\indent A.N. would like to acknowledge Vincent Beltrani for suggesting the use of the PSO algorithm. A.N. was supported by the Program in Plasma Science and Technology at Princeton University. We acknowledge support from the ARO and DOE.

\bibliographystyle{unsrt}
\bibliography{rabitzbib}

\begin{thebibliography}{10}

\bibitem{pulseshaping}
A.~Weiner.
\newblock {\em Rev. Sci. Instrum.}, 71(5), 2000.

\bibitem{lasers}
R.~S. Judson and H.~Rabitz.
\newblock {\em Phys. Rev. Lett.}, 68:1500--1503, 1992.

\bibitem{balint-kurti}
G.~Balint-Kurti, S.~Zou, and A.~Brown.
\newblock {\em Advances in Chemical Physics}, page~43.
\newblock Wiley, 2008.

\bibitem{W}
K.~W. Moore Tibbetts, C.~Brif, M.~D. Grace, A.~Donovan, D.~L. Hocker, T.~S. Ho, R.~B. Wu, and
  H.~Rabitz.
\newblock {\em Phys. Rev. A}, 86:062309, Dec 2012.

\bibitem{Moorewithcontrolstuff}
K.~W. Moore and H.~Rabitz.
\newblock {\em Phys. Rev. A}, 84:012109, 2011.

\bibitem{surveypaper}
R.~Chakrabarti and H.~Rabitz.
\newblock {\em Int. Rev. Phys. Chem.}, 26(4), 2007.

\bibitem{topologyscience}
H.~Rabitz, M.~Hsieh, and C.~Rosenthal.
\newblock {\em Science}, 303(5666):1998--2001, 2004.

\bibitem{kosut}
H.~Rabitz, T.~S. Ho, M.~Hsieh, R.~Kosut, and M.~Demiralp.
\newblock {\em Phys. Rev. A}, 74:012721, 2006.

\bibitem{roslund}
J.~Roslund and H.~Rabitz.
\newblock {\em Phys. Rev. A}, 80:013408, 2009.

\bibitem{complexity}
K.~W. Moore, M.~Hsieh, and H.~Rabitz.
\newblock {\em J. Chem. Phys.}, 128(15), 2008.

\bibitem{hamiltonianstructure}
A.~Donovan, V.~Beltrani, and H.~Rabitz.
\newblock {\em Phys. Chem. Chem. Phys.}, 13:7348--7362, 2011.

\bibitem{DMORPH}
A.~Rothman, T.~Ho, and H.~Rabitz.
\newblock {\em Phys. Rev. A}, 72(2), 2005.

\bibitem{otherDMORPH}
A.~Rothman, T.~Ho, and H.~Rabitz.
\newblock {\em J. Chem. Phys.}, 123(13), 2005.

\bibitem{PSO}
J.~Kennedy and R.~Eberhart.
\newblock In {\em Proc. IEEE Symposium on Swarm Intelligence}, pages
  1942--1948, 1995.

\bibitem{popdynam1}
Q.~Ren, G.~Balint-Kurti, F.~Manby, M.~Artamonov, T.~Ho, and H.~Rabitz.
\newblock {\em J. Chem. Phys.}, {124}({1}), {2006}.

\bibitem{photonicreagent}
V.~Beltrani, J.~Dominy, T.~Ho, and H.~Rabitz.
\newblock {\em J. Chem. Phys.}, 126(9), 2007.

\bibitem{Nanduri2}
A.~Nanduri, A.~Donovan, T.~Ho, and H.~Rabitz.
\newblock In preparation.
\newblock 2013.

\bibitem{constrained}
K.~W. Moore and H.~Rabitz.
\newblock {\em J. Chem. Phys.}, 137(13), 2012.

\end{thebibliography}

\newpage
\section*{Figures}

\end{document}